\documentstyle[11pt,cs10,html,epsf]{article}

\def\mj{$\,{\rm M}_{\rm J}\,$}
\def\mjj{{\rm M}_{\rm J}\,}
\def\etal{{\it et~al.}}
\def\lo{$L_\odot\,$}
\def\mo{$M_\odot\,$}

\def\star{*}
\def\rj{R$_{\rm J}\ $}

\def\etal{{\it et~al.\,}}
\def\mic{$\mu$m}
\def\sles{\lower2pt\hbox{$\buildrel {\scriptstyle <}
   \over {\scriptstyle\sim}$}}
\def\sgreat{\lower2pt\hbox{$\buildrel {\scriptstyle >}
   \over {\scriptstyle\sim}$}}

\begin{document}

\title{The Spectral Character of Giant Planets and Brown Dwarfs}
 
\author{A. Burrows\altaffilmark{1}, M. Marley\altaffilmark{2}, W.B. Hubbard\altaffilmark{3},
        D. Sudarsky\altaffilmark{1}, and C. Sharp\altaffilmark{1},
        J.I. Lunine\altaffilmark{3}, T. Guillot\altaffilmark{4}, D. Saumon\altaffilmark{5},
        R. Freedman\altaffilmark{6}}
 
\altaffiltext{1}{Department of Astronomy and Steward Observatory,
                 University of Arizona, Tucson, AZ \ 85721}
\altaffiltext{2}{Department of Astronomy, New Mexico State University,
                 Box 30001/Dept. 4500, Las Cruces NM 88003}
\altaffiltext{3}{Lunar and Planetary Laboratory, University of Arizona,
                 Tucson, AZ \ 85721}
\altaffiltext{4}{Department of Meteorology, University of Reading, P.O. Box 239,
                 Whiteknights, Reading RG6 6AU, United Kingdom}
\altaffiltext{5}{Department of Physics and Astronomy, Vanderbilt University, Nashville, TN 37235}
\altaffiltext{6}{Space Physics Research Institute, NASA Ames Research Center, Moffett Field CA 94035}

\begin{abstract}

Since October of 1995, employing Doppler spectroscopy astronomers have discovered as many as 20
giant planets and brown dwarfs, including companions to $\tau$ Boo, 51 Peg, $\upsilon$ And,
55 Cnc, $\rho$ CrB, 70 Vir, 16 Cyg B, and 47 UMa.
These discoveries
have galvanized the planetary science community, astronomers, and the public at large.
Within hours of the announcement of the planet 51 Peg b, the first {\it direct}
detection of an unimpeachable brown dwarf,
Gl229 B, was also announced.
Gl229 B is a watershed since it has methane spectral features
and a surface temperature below 1000 Kelvin, characteristics unique to objects
with substellar masses.
 
During the last two years, building upon our previous experience in the modeling of brown dwarfs and M stars,
we published theoretical studies of the evolution and spectra of extrasolar giant planets.
We have recently upgraded our capabilities and now generate
{\it non--gray} spectral and color models of both giant planets and brown dwarfs.
This theory will soon encompass objects whose effective temperatures range from 100 K to 4000 K
and whose masses span three orders of magnitude.
The evolutionary, spectral, and color calculations upon which we have embarked
are in direct support of the searches
now being planned in earnest
with the HST (WFPC2, NICMOS), the IRTF, the MMT 6.5-meter upgrade, the Large Binocular Telescope (LBT),
Keck's I and II, ISO, UKIRT, NGST, the VLT, COROT, DENIS, 2MASS,
and SIRTF.

\end{abstract}


\section{Introduction}

Using Doppler spectroscopy astronomers are populating a zoo of 
giant planets and brown dwarfs orbiting nearby stars,
including $\tau$ Boo, 51 Peg, $\upsilon$ And,
55 Cnc, $\rho$ CrB, 70 Vir, 16 Cyg B, and 47 UMa (Noyes \etal\ 1997; Butler \etal\ 1997; Cochran \etal\ 1997;
Marcy \& Butler 1996; Butler \& Marcy 1996; Mayor \& Queloz 1995;
Latham \etal\ 1989). 
Table 1 lists these newly--discovered planets/brown dwarfs, in order of increasing semi--major axis, along
with the giant planets in our solar system and the brown dwarf Gl229 B.
The direct detection of Gl229 B
(Oppenheimer \etal\  1995; Nakajima \etal\ 1995; Matthews \etal\ 1996; Geballe \etal\ 1996;
Marley \etal\ 1996; Allard \etal\ 1996; Tsuji \etal\ 1996) was a watershed,
since it has methane spectral features
and an inferred effective temperature (in this case, T$_{\rm eff}$$\sim$950 K) far
below that at the solar--metallicity main sequence edge ($\sim$1750 K, Burrows \etal\ 1993).
In addition, the almost complete absence of the spectral signatures of metal oxides and hydrides (such as
TiO, VO, FeH, and CaH) is in keeping with theoretical predictions that these species are depleted
in the atmospheres of all but the youngest (hence, hottest) substellar objects
and are sequestered in condensed form below the photosphere (Lunine \etal\ 1989; Marley \etal\ 1996).
Also shown in Table 1 are the primary's metallicity
(G. Gonzalez, private communication), \nobreak{M$_{\rm p}$sin (i)}, orbit period,
eccentricity, and distance to the sun. 
The wide range in mass
and period, as well as the proximity of many of the planets/brown dwarfs
to their primaries, was not anticipated by most planetary scientists.
Though the technique of Doppler spectroscopy used to find most of these companions
selects for massive, nearby
objects, their variety and existence is a challenge to conventional
theory.
Since direct detection is now feasible, and has been
demonstrated by the recent acquisition of Gl229 B, it is crucial for the future of extrasolar planet
searches that the spectra, colors, evolution, and physical structure of objects from Saturn's mass (0.3$\times$
Jupiter's mass, \mj) to 70 \mj
be theoretically investigated.

In parallel with these recent planetary and brown dwarf discoveries has been the much
more detailed scrutiny in the optical and the near--infrared
of hot young brown dwarfs and late M dwarfs near the
main--sequence edge ({\it cf.} Kirkpatrick, Henry, \& Simons 1995; Jones \& Tsuji 1997;
Zapatero-Osorio \etal\ 1996; Tinney \etal\ 1995; Delfosse 1997).  Finally, we are
obtaining photometry and spectroscopy for collections of objects that {\it bracket} the substellar limit.
What they reveal is a plethora of objects that populate a rising mass function, though not one that rises fast
enough to have significant dynamical consequences for the galactic disk or halo (Chabrier \& Baraffe 1997).
Nevertheless, it is clear that edge objects reside in an unanticipated region of the H--R diagram and are much
bluer
in the near infrared than one would have expected from naive extrapolations of the early M dwarf sequence.
Grain formation, metal depletions ({\it e.g.}, TiO into perovskite, CaTiO$_3$),
and a shift of the important molecules with decreasing T$_{\rm eff}$ are needed
to understand high--gravity atmospheres from 3000 K to 1500 K
(Lunine \etal\ 1989; Tsuji \etal\ 1996; Allard \etal\ 1996).  Clearly, new theoretical insights are required to
 separate
very young and hot brown dwarfs from the more massive M dwarfs.

During the last two years, building upon our previous experience in the modeling of brown dwarfs and M stars,
we published theoretical studies of the evolution and spectra of extrasolar giant planets (EGPs)
(Burrows \etal\ 1995; Saumon \etal\ 1996; Guillot \etal\ 1996; Marley \etal\ 1996; Burrows \etal\ 1997).
This work, initiated before the first discovery announcements in October of 1995, is described below.
 
Some of the space platforms and new ground--based facilities that have or will
obtain infrared and optical data of extrasolar planets and brown dwarfs include
the HST (WFPC2, NICMOS), the IRTF, the MMT 6.5-meter upgrade (Angel 1994), the Large Binocular Telescope (LBT)
(planned for Mt. Graham), Keck's I and II, the European ISO, UKIRT, NGST, the VLT, COROT, DENIS, 2MASS,
and SIRTF.
One project of the Keck I and II facility,
under the aegis of NASA's ASEPS-0 (Astronomical Study of Extrasolar Planetary Systems)
program, will be to search for giant planets around nearby stars.
A major motivation for the Palomar Testbed Interferometer supported by NASA
is the search for extrasolar planets.
Recently, the NASA administrator outlined a program to detect planetary
systems around nearby stars that may become a future scientific focus of NASA.
This vision is laid out in the {\it Exploration of Neighboring Planetary Systems (ExNPS)
Roadmap} (see also the ``TOPS'' Report, 1992).

 \begin{table}
 \caption{\bf The Bestiary}
 \begin{center}\scriptsize
 \begin{tabular}{cccccccccc}
Object&Star&M$_\star$ (M$_\odot$)&L$_\star$ (L$_\odot$)&
d (pc)&[Fe/H]&M (M$_J$)&$a$ (AU)&P (days)&
$e$\\
 \tableline
 HD283750&K2V?&0.75?&0.2?&16.5&?&$\sgreat$50&$\sim$0.04&1.79&0.02\\
 $\tau$ Boo b&F7V&1.25&2.5&15&+0.34&$\sgreat$3.44&0.046&3.313&0.0162\\
 51 Peg b&G2.5V&1.0&1.0&15.4&+0.21&$\sgreat$0.45&0.05&4.23&0.0\\
 HD98230&G0V&1.1&1.5&7.3&-0.12&$\sgreat$37&0.05&3.98&0.0\\
 $\upsilon$ And b&F7V&1.25&2.5&17.6&+0.17&$\sgreat$0.68&0.058&4.61&0.109\\
 55 Cnc b&G8V&0.85&0.5&13.4&+0.29&$\sgreat$0.84&0.11&14.76&0.051\\
 $\rho$ CrB b&G0V&1.0&1.77&17.4&-0.19&$\sgreat$1.13&0.23&39.65&0.028\\
 HD112758&K0V&0.8&0.4&16.5&?&$\sgreat$35&$\sim$0.35&103.22&0.16\\
 HD114762&F9V&1.15&1.8&28&-0.60&$\sgreat$10&0.38&84&0.25\\
 70 Vir b&G4V&0.95&0.8&18.1&-0.03&$\sgreat$6.9&0.45&116.7&0.40\\
 HD140913&G0V&1.1&1.5&?&?&$\sgreat$46&$\sim$0.54&147.94&0.61\\
 HD89707&G1V&1.1&1.3&24.5&-0.423&$\sgreat$54&0.69&198.25&0.95\\
 BD -04782&K5V&0.7&0.1&?&?&$\sgreat$21&$\sim$0.7&240.92&0.28\\
 HD110833&K3V&0.75&0.2&16&?&$\sgreat$17&$\sim$0.8&270.04&0.69\\
 HD217580&K4V&0.7&0.1&18.5&?&$\sgreat$60&$\sim$1.0&1.24 yrs&0.52\\
 HD18445&K2V&0.75&0.2&$\sim$20&?&$\sgreat$39&1.2&1.52 yrs&0.54\\
 16 Cyg Bb&G2.5V&1.0&1.0&22&0.11&$\sgreat$1.66&1.7&2.19 yrs&0.57\\
 47 UMa b&G0V&1.1&1.5&14.1&+0.01&$\sgreat$2.4&2.1&2.98 yrs&0.03\\
 HD29587&G2V&1.0&1.0&42&?&$\sgreat$40&2.1&3.17 yrs&0.0\\
 Gl 411 b&M2V&0.4&0.02&2.52&-1.0&$\sgreat$0.9&2.38&5.8 yrs&?\\
 55 Cnc c&G8V&0.85&0.5&13.4&+0.29&$\sgreat$5&3.8& $>$8 yrs&?\\
 Jupiter&G2V&1.0&1.0&0.0&0.0&1.00&5.2&11.86 yrs&0.048\\
 Saturn&G2V&1.0&1.0&0.0&0.0&0.3&9.54&29.46 yrs&0.056\\
 Gl 229 B&M1V&0.45&0.03&5.7&+0.20&30-55&$\sgreat$44.0&$\sgreat$400 yrs&?\\
 \end{tabular}
 \end{center}
 \end{table}

\section{Previous Gray Models of Giant Planets and Brown Dwarfs}
 
In 1995, our group (Burrows, Saumon, Guillot, Hubbard \& Lunine 1995;
Saumon, Hubbard, Burrows, Guillot, Lunine, \& Chabrier 1996) calculated
a suite of models of the evolution and emissions
of EGPs, under the problematic black body assumption.
We derived fluxes and dimensions as a function of age, composition,
and mass, both as a guide for giant planet searches
and as a tool for interpreting the results of any
positive detections.  It has long been recognized (D'Antona \& Mazzitelli 1985; Stevenson 1991) that
the same physics governs the structure and evolution
of the suite of electron-degenerate and hydrogen-rich objects
ranging from M dwarfs (\sgreat 0.08 \mo) and brown dwarfs (at the high--mass end) to Jupiters and
Saturns (at the low--mass end).
Surprisingly,
no one had
accurately mapped out the properties of
objects between the mass of giant planets in our solar system and the traditional
brown dwarfs (\sgreat 15 $\mjj$, where $\mjj$ is the mass of
Jupiter, $\sim$ 10$^{-3}$ \mo).  This is precisely the mass range for
many of the newly--discovered planets listed in Table 1.

EGPs radiate in the optical by reflection and in the
infrared by the thermal emission of both absorbed stellar light and the planet's
own internal energy.
To calculate their cooling curves, we used the Henyey code previously constructed to study
brown dwarfs and M dwarfs (Burrows, Hubbard, \& Lunine 1989; Burrows \etal\ 1993).
Below effective temperatures (T$_{\rm eff}$) of 600 K, we employed the atmospheres
of Graboske \etal\ (1975), who included opacities due to water, methane, ammonia,
and collision-induced absorption by H$_{2}$ and He.  The gravity dependence of the EGP atmospheres was
handled as in Hubbard (1977).  Above T$_{\rm eff}=600\,$K,  we used the X model of Burrows \etal\ (1993).
The two prescriptions were interpolated in the overlap region. We employed the hydrogen/helium equation of
state of Saumon \& Chabrier (1991, 1992) and Saumon, Chabrier, \& Van Horn (1995)
and ignored rotation
and the possible presence of an ice/rock core (Pollack 1984; Bodenheimer \& Pollack 1986).
In Burrows \etal\ (1995) and Saumon \etal\ (1996), the EGPs were assumed to be fully convective at all times.
We included the effects of ``insolation'' by a central star and considered
semi-major axes ($a$) between 2.5 A.U. and 20 A.U.  Giant
planets may form preferentially near 5 A.U. (Boss 1995), but as the new data dramatically affirm,
a broad range of $a$s can not be excluded.
We assumed that the Bond albedo of an EGP is that of Jupiter (0.35, Conrath, Hanel, \& Samuelson 1989).
For the Burrows \etal\ (1995) study, we evolved EGPs with masses from 0.3 \mj $\,$(the mass of Saturn)
through 15 \mj .
Whether a 15 \mj object is a planet or a brown dwarf is largely a semantic issue, though one might
distinguish gas giants and brown dwarfs by their
mode of formation ({\it e.g.}, in a disk or ``directly'').  Physically, compact hydrogen-rich objects with mass
es from 0.00025 \mo$\,$ through
0.25 \mo$\,$ form a continuum.  However, EGPs
above $\sim$13 \mj$\,$ do burn ``primordial'' deuterium for up to 10$^{8}$ years.

If 51 Peg b is a gas giant, its radius is only 1.2$\,$R$_{\rm J}$ and its
luminosity is about $3.5\times 10^{-5}\,L_\odot$.  This bolometric luminosity is more than $1.5 \times 10^4$
times the present luminosity of Jupiter and only a factor of two below that at the edge of the main sequence.
The radiative region encompasses
the outer 0.03\% in mass, and 3.5\% in radius.
The study by Guillot \etal\ (1996) demonstrated that 51 Peg b is well within its Roche lobe
and is not experiencing significant photoevaporation.  Its deep potential well ensures that,
even so close to its parent, 51 Peg b is {\bf stable}.
If 51 Peg b were formed beyond an A.U. and moved inward on a timescale greater than $\sim 10^{8}$ years, it 
would closely follow the $R_{\rm p} \sim$ \rj trajectory to its equilibrium position.

\begin{figure}
\plotfiddle{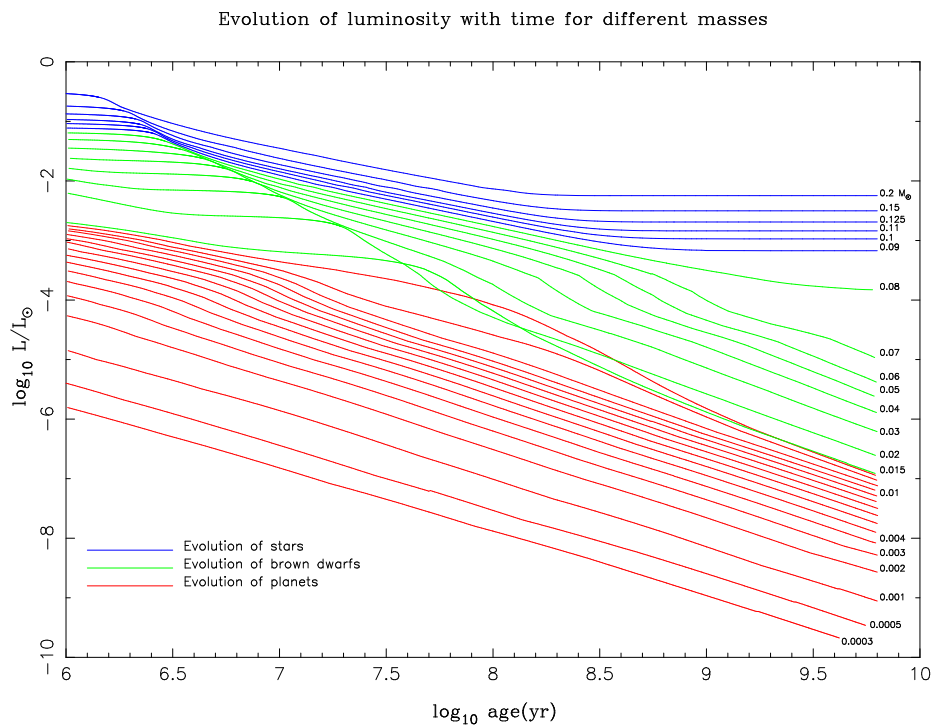}{3truein}{-90}{50}{50}{-200}{+300}
\caption{Evolution of the luminosity (in L${_\odot}$) of solar--metallicity M dwarfs and substellar objects
versus time (in years) after formation.
The masses in M${_\odot}$ label most of the curves, with the lowest three curves
corresponding to a ``Saturn,'' an EGP with half the mass of Jupiter, and a ``Jupiter,''
and the upper curve corresponding to a 0.2 \mo M dwarf.  The solid lines at higher luminosities are for M 
dwarfs, those
at lower luminosities are for EGPs that burn little or no deuterium, and the dotted lines are for brown dwarfs
or transition objects that burn deuterium, but don't settle onto the main sequence within
10$^{10}$ years
(from Burrows \etal\ 1997).
}
\end{figure}

\section{New Non--Gray Models}
 
However, to credibly estimate the infrared band fluxes and improve upon the black body assumption made in
Burrows \etal\ (1995)
and  Saumon \etal\ (1996), we have
recently performed {\bf non-gray} simulations at solar--metallicity of the evolution, spectra,
and colors of isolated EGP/brown dwarfs down to T$_{\rm eff}$s of 100 K (Burrows \etal\ 1997).
Figure 1 portrays the luminosity versus time for objects from Saturn's mass (0.3 \mj) to 0.2 \mo for this model
suite.

The early plateaux between 10$^6$ years and 10$^8$ years are due to deuterium burning, where
the initial deuterium mass fraction was taken to be 2$\times$10$^{-5}$.  Deuterium burning occurs earlier,
is quicker, and is at higher luminosity for the more massive models, but can take as long
as 10$^{8}$ years for a 15 \mj object.  The mass below which less than 50\% of the ``primordial''
deuterium is burnt is $\sim$13 \mj (Burrows \etal\ 1995).  On this figure, we have arbitrarily
classed as ``planets'' those objects that do not burn deuterium and as ``brown dwarfs'' those that do burn 
deuterium, but not light hydrogen.  While this distinction is physically motivated, we do not
advocate abandoning the definition based on origin.  Nevertheless, the separation
into M dwarfs, ``brown dwarfs'', and giant ``planets'' is useful for parsing by eye 
the information in the figure
 
In Figure 1, the bumps between 10$^{-4}$ \lo and 10$^{-3}$ \lo and between 10$^{8}$ and 10$^{9}$ years,
seen on the cooling curves of objects from 0.03 \mo to 0.08 \mo, are due to silicate and iron grain formation.
These effects,
first pointed out by Lunine \etal\ (1989), occur for T$_{\rm eff}$s between 2500 K and 1300 K.
The presence of grains affects the precise mass and luminosity at the edge of the main sequence.
Since grain and cloud models are problematic, there still remains much to learn concerning
their role and how to model them (Lunine \etal\ 1989; Allard \etal\ 1997).

To constrain the properties of the brown dwarf Gl229 B (Oppenheimer \etal\ 1995;
Nakajima \etal\ 1995; Geballe \etal\ 1996; Matthews \etal\ 1996), we (Marley, Saumon, Guillot,
Freeman, Hubbard, Burrows, \& Lunine 1996)
constructed a grid of
atmospheres with T$_{\rm eff}$ ranging from 600 to 1200 K and
$1.0\times10^4$ cm s$^{-2}$ $< {\rm gravity} <$ 3.2$\times10^5$ cm s$^{-2}$.  For each case we computed a 
self--consistent
radiative-convective equilibrium temperature profile and the emergent radiative flux.
By comparing our theoretical spectra with the UKIRT (Geballe \etal\ 1996) and
HST (Matthews \etal\ 1996) data, we derived an effective temperature of
$960 \pm 70$ K and a gravity between $0.8 \times 10^{5}$ cm s$^{-2}$ and $2.2 \times 10^{5}$ cm s$^{-2}$.
These results translate into masses and ages of 20--55 \mj and 0.5--5 Gyr, respectively.
Gravity maps almost directly
into mass, and ambiguity in the former results in uncertainty in the latter.
The Marley \etal\ (1996) study was nicely complemented in the literature by
the Gl229 B calculations of Allard \etal\ (1996) and Tsuji \etal\ (1996).

The studies of Burrows \etal\ (1997) and Marley \etal\ (1996) revealed major new aspects of
EGP/brown dwarf atmospheres that bear listing and that uniquely
characterize them.
Below T$_{\rm eff}$s of 1300 K, the dominant equilibrium carbon molecule is CH$_4$, not CO,
and below 600 K the dominant nitrogen molecule is NH$_3$, not N$_2$ (Fegley \& Lodders 1996).
The major opacity sources are H$_2$, H$_2$O, CH$_4$, and NH$_3$.
For T$_{\rm eff}$s below $\sim$400 K, water clouds form at or above the photosphere
and for T$_{\rm eff}$s below 200 K, ammonia clouds form ({\it viz.,} Jupiter).  Collision--induced absorption
of H$_2$ partially suppresses emissions longward of $\sim$10 \mic.  The holes in the opacity
spectrum of H$_2$O that define the classic telluric IR bands also regulate much of the emission from
EGP/brown dwarfs in the near infrared.  Importantly, the windows in H$_2$O and the suppression by H$_2$ 
conspire to force flux to the blue for a given T$_{\rm eff}$.
The upshot is an exotic spectrum enhanced relative to the black body value
in the $J$ and $H$ bands ($\sim$1.2 \mic\ and $\sim$1.6 \mic, respectively) by as much as {\it two} to {\it ten}
orders of magnitude,
depending upon T$_{\rm eff}$.
Figure 2 depicts spectra between 1 \mic\ and 10 \mic\ at a detector 10 parsecs
away from solar--metallicity objects with age 1 Gyr and masses
from 1 \mj through 40 \mj.

\begin{figure}
\plotfiddle{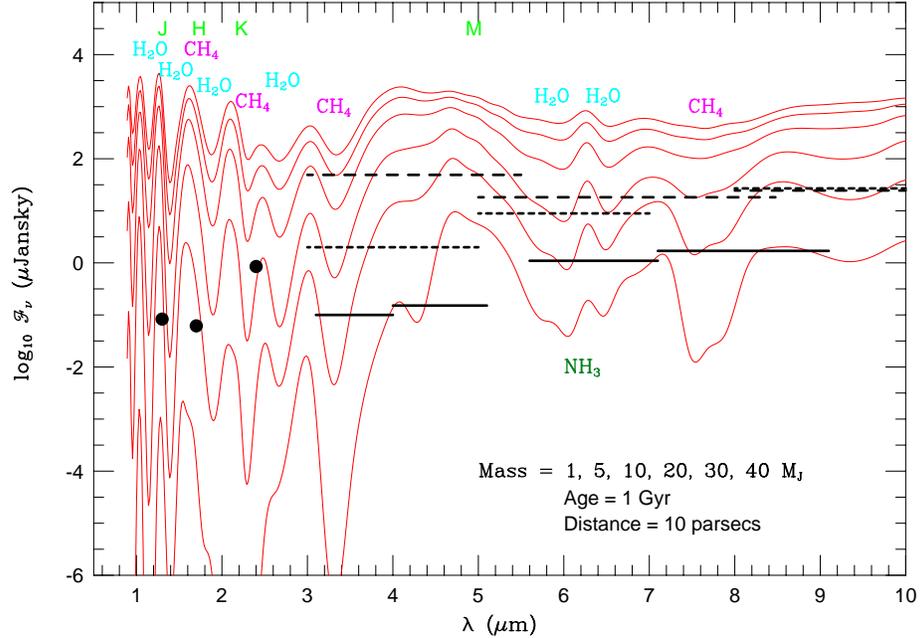}{3truein}{-270}{50}{50}{185}{-30}
\caption{The flux (in $\mu$Janskys) at 10 parsecs versus wavelength (in microns) from 1 \mic\ to
10 \mic\ for 1, 5, 10, 20, 30, and 40\mj solar--metallicity models at 1 Gyr.   Shown are the positions of the
$J$,$H$, $K$, and $M$ bands and various molecular absorption features.  Superposed for comparison
are the putative sensitivities of the three NICMOS cameras, ISO,
Gemini/SOFIA, and SIRTF.  NICMOS is denote with large black dots, ISO with long dashes,
Gemini/SOFIA with short dashes, and SIRTF with solid lines.
At all wavelengths, SIRTF's projected sensitivity is greater than ISO's.
SOFIA's sensitivity overlaps with that of ISO around 10\mic.  For other wavelength intervals, the order
of sensitivity is SIRTF $>$ Gemini/SOFIA $>$ ISO, where $>$ means ``is more sensitive than''
(from Burrows \etal\ 1997).
}
\end{figure}

Superposed are putative sensitivities
for the three NICMOS cameras (Thompson 1992), ISO (Benvenuti \etal\ 1994), SIRTF (Erickson \& Werner 1992),
and Gemini/SOFIA (Mountain \etal\ 1992; Erickson 1992).
Figure 2 demonstrates how unlike a black body an EGP spectrum
is. For example, the enhancement at 5 \mic\ for a 1 Gyr old, 1 \mj extrasolar planet is by four orders
of magnitude.
Figure 3 portrays the evolution from 0.1 Gyr to 5 Gyr of the spectrum from 1 \mic\ to 10 \mic\
of a 10\mj object in isolation, without a reflected component.   The higher curves are for the younger ages.
and some of the molecular features are identified.

\begin{figure}
\plotfiddle{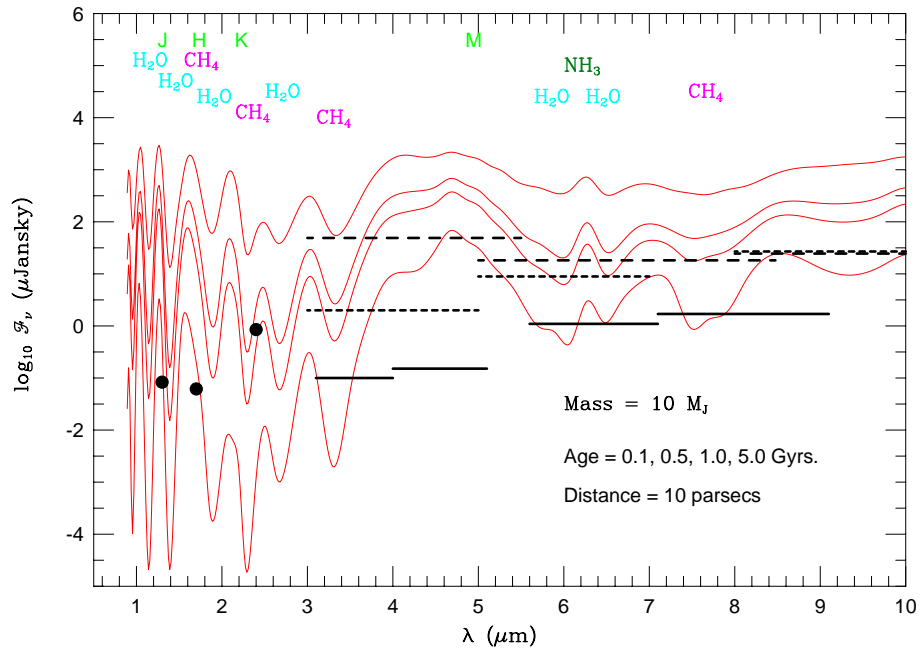}{3truein}{-270}{50}{50}{185}{-30}
\caption{The flux (in $\mu$Janskys) at 10 parsecs versus wavelength (in microns) from 1 \mic\ to
10 \mic\ for a 10 \mj solar--metallicity object at ages of 0.1, 0.5, 1.0, and 5.0 Gyr.
Superposed are the positions of the $J$, $H$, $K$, and $M$ bands,
the estimated sensitivities of the three NICMOS cameras, ISO, Gemini/SOFIA,
and SIRTF (see Figure 2), and the positions of various of the
important molecular absorption features (from Burrows \etal\ 1997).
}
\end{figure}

As T$_{\rm eff}$ decreases below $\sim$1000 K, the flux in the $M$ band ($\sim$5 \mic)
is progressively enhanced relative to the black body value.
While at 1000 K there is no enhancement, at 200 K it is near 10$^5$.   Hence, the $J$, $H$, and $M$ bands
are the premier bands in which to search for cold substellar objects.   The $Z$ band ($\sim$1.05 \mic) is also
super--black--body over this T$_{\rm eff}$ range.  However, there is a NH$_3$ feature
in the $Z$ band that was not in our database.  This
will likely reduce the flux in this band for the cooler models.
Eventhough $K$ band ($\sim$2.2 \mic) fluxes are generally higher
than black body values, H$_2$ and CH$_4$ absorption features in the $K$ band decrease its importance
{\it relative} to $J$ and $H$.  As a consequence of the increase of atmospheric
pressure with decreasing T$_{\rm eff}$, the anomalously blue $J-K$ and $H-K$
colors get {\it bluer},
not redder.

\begin{figure}
\plotone{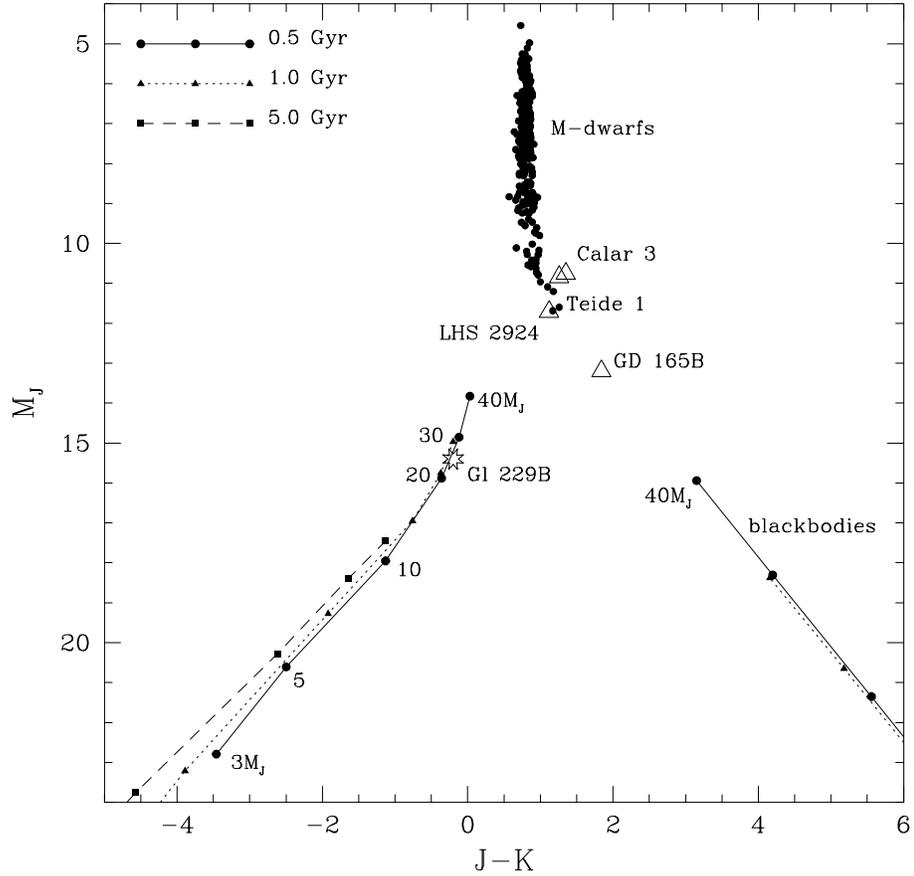}
\caption{Absolute $J$ vs. $J-K$ color--magnitude diagram at solar metallicity.  Theoretical isochrones
are shown for $t$ = 0.5, 1, and 5 Gyr, along with their black body
counterparts.  The difference between black body colors and model
colors is striking.  The brown dwarf, Gliese 229 B (Oppenheimer \etal\ 1995),
the young brown dwarf candidates Calar 3 and Teide 1 (Martin, Rebolo, \& Zapatero-Osorio 1996),
and late M dwarfs LHS 2924 and
GD165B (Kirkpatrick, Henry, \& Simons 1994,1995)) are plotted for comparison.
The lower main sequence is defined by a selection of M--dwarf stars from
Leggett (1992) (from Burrows \etal\ 1997).
}
\end{figure}

To illustrate this, Figure 4 shows a representative color--magnitude diagram ($J$ versus $J-K$)
for solar--metallicity objects with masses from 3 \mj to 40 \mj, for ages of 0.5, 1.0, and 5.0 Gyr.
Included are the corresponding black body curves,
hot, young brown dwarf or extremely late M dwarf
candidates such
as LHS2924, GD 165B, Calar 3, and Teide 1 (Kirkpatrick, Henry, \& Simons 1994,1995; Zapatero-Osorio, Martin, \&
Rebolo 1997;
Martin, Rebolo, \& Zapatero-Osorio 1996),
and a sample of M dwarfs from Leggett (1992).
This figure illustrates the unique color realms occupied by extrasolar giant planets and brown dwarfs.
Figure 4 portrays the fact that the $K$ and $J$ versus $J-K$ infrared H--R diagrams loop back to the blue
below the edge
of the main sequence and are not continuations of the M dwarf sequence into the red.
The difference between the black body curves and the model curves is between 3 and 10 magnitudes
for $J$ versus $J-K$, more for $K$ versus $J-K$.  Gl229 B fits nicely among these theoretical isochrones.
The suppression of $K$ by H$_2$ and CH$_4$ features is largely responsible for this anomalous blueward
trend with decreasing mass and T$_{\rm eff}$.
Tables 2 and 3 depict absolute magnitudes and colors in the IR as a function of gravity and
effective temperature for the full range of EGPs and brown dwarfs.  The gravity dependences of the colors
are not small.  However, these tables should be viewed with caution until higher--resolution spectra
and a better methane opacity database are both available. 

 \begin{table}
 \caption{Absolute Magnitudes of Synthetic BD/EGPs, [M/H]=0.0 \tablenotemark{\dag}}
 \begin{center}\scriptsize
 \begin{tabular}{cccccccc}
$g$\ (cm s$^{-2}$) & $T_{eff}$ (K) & $M_J$ & $M_H$ & $M_K$ & $M_{L^{\prime}}$ & $M_M$ & $M_N$ \\
 \tableline
$10^5$&$  1000.0$&$   15.35$&$   14.99$&$   15.62$&$   13.34$&$   12.56$&$   12.51$ \\
&$   800.0$&$   16.44$&$   16.01$&$   17.09$&$   14.13$&$   13.14$&$   13.29$ \\
&$   600.0$&$   17.91$&$   17.40$&$   19.27$&$   15.18$&$   13.94$&$   14.21$ \\
&$   500.0$&$   18.96$&$   18.39$&$   20.96$&$   15.94$&$   14.49$&$   14.77$ \\
$3\times10^4$&$  1000.0$&$   14.96$&$   14.52$&$   14.98$&$   12.99$&$   12.12$&$   12.03$ \\
&$   800.0$&$   16.04$&$   15.54$&$   16.40$&$   13.83$&$   12.71$&$   12.83$ \\
&$   600.0$&$   17.49$&$   16.92$&$   18.43$&$   14.94$&$   13.51$&$   13.78$ \\
&$   400.0$&$   20.33$&$   19.69$&$   22.97$&$   16.97$&$   14.96$&$   15.04$ \\
&$   300.0$&$   22.62$&$   21.91$&$   26.74$&$   18.58$&$   16.08$&$   15.74$ \\
$10^4$&$  1000.0$&$   14.66$&$   14.11$&$   14.26$&$   12.70$&$   11.70$&$   11.57$ \\
&$   800.0$&$   15.82$&$   15.23$&$   15.88$&$   13.66$&$   12.39$&$   12.49$ \\
&$   600.0$&$   17.27$&$   16.61$&$   17.89$&$   14.84$&$   13.24$&$   13.50$ \\
&$   400.0$&$   19.76$&$   19.31$&$   21.86$&$   16.74$&$   14.64$&$   14.71$ \\
$3\times10^3$&$  1000.0$&$   13.97$&$   13.15$&$   13.05$&$   12.31$&$   11.16$&$   10.95$ \\
&$   800.0$&$   15.42$&$   14.64$&$   15.00$&$   13.38$&$   11.97$&$   12.07$ \\
&$   600.0$&$   17.17$&$   16.48$&$   17.44$&$   14.72$&$   12.97$&$   13.21$ \\
&$   400.0$&$   19.84$&$   19.40$&$   21.49$&$   16.77$&$   14.49$&$   14.54$ \\
&$   200.0$&$   24.60$&$   24.03$&$   28.84$&$   19.93$&$   16.58$&$   15.91$ \\
 \end{tabular}
 \end{center}
 \tablenotetext{\dag}{We employed the transmission curves of Bessell \& Brett (1988) and
Bessell (1990) to define the photometric bandpasses
and the model of Vega by Dreiling \& Bell (1980) for the calibration
of the magnitude scale. Table taken from Burrows \etal\ 1997.}
 \end{table}

 \begin{table}
 \caption{Color Indices of Synthetic BD/EGPs, [M/H]=0.0 \tablenotemark{\dag}}
 \begin{center}\scriptsize
 \begin{tabular}{ccccccc}
$g$\ (cm s$^{-2}$) & $T_{eff}$ (K)&$J-H$&$J-K$&$H-K$&$K-L^{\prime}$&$M-N$ \\
 \tableline
$10^5$&  1000.0&    0.35&   -0.28&   -0.63&    2.28&    0.05  \\
&   800.0&    0.43&   -0.65&   -1.08&    2.96&   -0.15  \\
&   600.0&    0.51&   -1.36&   -1.87&    4.09&   -0.27  \\
&   500.0&    0.57&   -2.00&   -2.57&    5.02&   -0.27  \\
$3\times10^4$&  1000.0&    0.44&   -0.02&   -0.46&    1.99&    0.09  \\
&   800.0&    0.50&   -0.37&   -0.87&    2.58&   -0.12  \\
&   600.0&    0.57&   -0.94&   -1.51&    3.49&   -0.27  \\
&   400.0&    0.63&   -2.64&   -3.27&    5.99&   -0.08  \\
&   300.0&    0.71&   -4.12&   -4.83&    8.17&    0.34  \\
$10^4$&  1000.0&    0.55&    0.40&   -0.15&    1.55&    0.13  \\
&   800.0&    0.59&   -0.06&   -0.65&    2.22&   -0.10  \\
&   600.0&    0.66&   -0.62&   -1.28&    3.05&   -0.26  \\
&   400.0&    0.46&   -2.10&   -2.56&    5.12&   -0.08  \\
$3\times10^3$&  1000.0&    0.82&    0.92&    0.10&    0.74&    0.21  \\
&   800.0&    0.78&    0.42&   -0.36&    1.62&   -0.10  \\
&   600.0&    0.69&   -0.28&   -0.97&    2.73&   -0.24  \\
&   400.0&    0.45&   -1.65&   -2.10&    4.73&   -0.05  \\
&   200.0&    0.56&   -4.25&   -4.81&    8.91&    0.67  \\
 \end{tabular}
 \end{center}
 \tablenotetext{\dag}{Taken from Burrows \etal\ 1997.}
 \end{table}

Ignoring for the moment the question of angular resolution,
one can compare the theoretical solar--metallicity spectrum and color
predictions of Burrows \etal\ (1997) with
putative detector sensitivities to derive encouraging detection ranges.
For example, at 5 \mic, SIRTF
might see a 1 Gyr old, 1 \mj object in isolation out to nearly 100 parsecs.   The range of NICMOS in $H$
for a 1 Gyr old, 5 \mj object is approximately 300 parsecs, while for a coeval 40 \mj object
it is near 1000 parsecs.  Furthermore, SIRTF will be able to see at 5 \mic\
a 5 Gyr old, 20 \mj object in isolation out to
$\sim$400 parsecs and NICMOS will be able to see at $J$ or $H$ a 0.1 Gyr old object with the same
mass out to $\sim$2000 parsecs.  These are dramatic numbers that serve to illustrate both the
promise of the new detectors and the enhancements we theoretically predict.

\section{Model Ingredients}

The ingredients needed to generate non--gray spectral, color, and evolutionary models
of M dwarfs, brown dwarfs, and EGPs are clear.  They include (1) equations of state for
metallic hydrogen/helium mixtures and molecular atmospheres, (2) chemical equilibrium
codes and thermodynamic data to determine the molecular fractions, (3) scattering and absorption opacities for
the dominant chemical species, (4) an atmosphere code to calculate temperature/pressure
profiles and to identify the radiative and convective zones, (5) an algorithm for converting a grid
of atmospheres into boundary conditions for evolutionary calculations,
(6) a Henyey code, and (7) a radiative transfer code to provide emergent
spectra.  In principle, the calculation of the atmosphere, involving as it does radiative transfer,
and the calculation of the emergent spectrum are done together.

\subsection{Opacities}

For solar metallicity,
near and above brown dwarf/EGP photospheres, throughout most of their lives
the dominant equilibrium form of carbon is CH$_4$, not CO, that
of oxygen is H$_2$O, and that of nitrogen is either N$_2$ or NH$_3$, depending upon T$_{\rm eff}$ (Fegley \& 
Lodders 1996).
Hydrogen is predominantly in the form
of H$_2$.  Silicates and metals are found at high optical depths and temperatures.
Clouds of NH$_3$ and H$_2$O can form for T$_{\rm eff}$s below
$\sim$200 K and  $\sim$400 K, respectively.  For the Burrows \etal\ (1997) models, we precipitated species
according
to their condensation curves, but did not consistently incorporate the effects of the associated
clouds.  If a species condensed, it was left at its saturated vapor pressure.
 
Water is an important source of opacity in late stellar and substellar objects, particularly when the
many lines that originate from highly excited energy levels are considered.
Depending upon the temperature of the layer, and the assumed
abundance of water, well over $2.0\times10^{8}$ lines could be required for a calculation
at the highest temperatures, while far fewer lines are needed at lower temperatures.
We use the new Partridge \& Schwenke H$_2$O database.
A synopsis of the opacity and line profile data we have employed to
simulate EGP and brown dwarf models follows.

\subsubsection{Line Lists:}

\begin{itemize}

\item Partridge \& Schwenke (1997) H$_2$O database: $3.0\times10^{8}$ lines 

\item HITRAN database (Rothman \etal\ 1992, 1997) 
\item GEISA database (Husson \etal\ 1997)
\item CH$_4$ and CH$_3$D, $1.9\times10^6$ lines; CO, $99,000$ lines, NH$_3$, $11,400$
lines, PH$_3$, $11,240$ lines, H$_2$S, $179,000$ lines (Tyuterev \etal\ 1994; Goorvitch 1994; 
Tipping 1990; Wattson \& Rothman 1992; L. R. Brown, private communication)

\item Modeled continuum opacity sources: 
$\rm H^-$
and $\rm H_2^-$ opacity
and collision--induced absorption (CIA)  
of H$_2$ and helium (Borysow \& Frommhold 1990; Zheng \& Borysow 1995); 
Rayleigh scattering: Rages \etal\ (1991)  

\end{itemize}

\subsubsection{Line Profiles:}
 
\begin{itemize}
 
\item Line widths
assumed to be due to H$_2$ or
H$_2$ + He broadening:
H$_2$O (Brown \& Plymate 1996; Gamache, Lynch, \& Brown 1996), CO (Bulanin \etal\ 1984; LeMoal \& Severin 1986);
CH$_4$ (Margolis 1993,1996; L. R. Brown, private communication); PH$_3$ (Levy, Lacome, \& Tarrago 1994);
NH$_3$ (Brown \& Peterson 1994)
\item For other species, use N$_2$-- and O$_2$--induced widths
 
\end{itemize}
 
\subsubsection{The k--coefficient Method:}
 
\begin{itemize}
 
\item The k--coefficient
method (Goody \etal\ 1989; Lacis \& Oinas 1991),  widely used in
planetary atmosphere and global climate modeling
\item Not the ODF technique (Saxner \& Gustafsson 1984).
That the gases have already
been mixed {\it before} the k coefficients are derived greatly diminishes
problems.
\item Typical errors in planetary atmospheres:
between 1 and 10\%
(Grossman \& Grant 1992,1994)
\item Supplemented by new Feautrier code, Bergeron atmosphere's code, {\it etc.} 
 
\end{itemize}

\begin{figure}
\plotfiddle{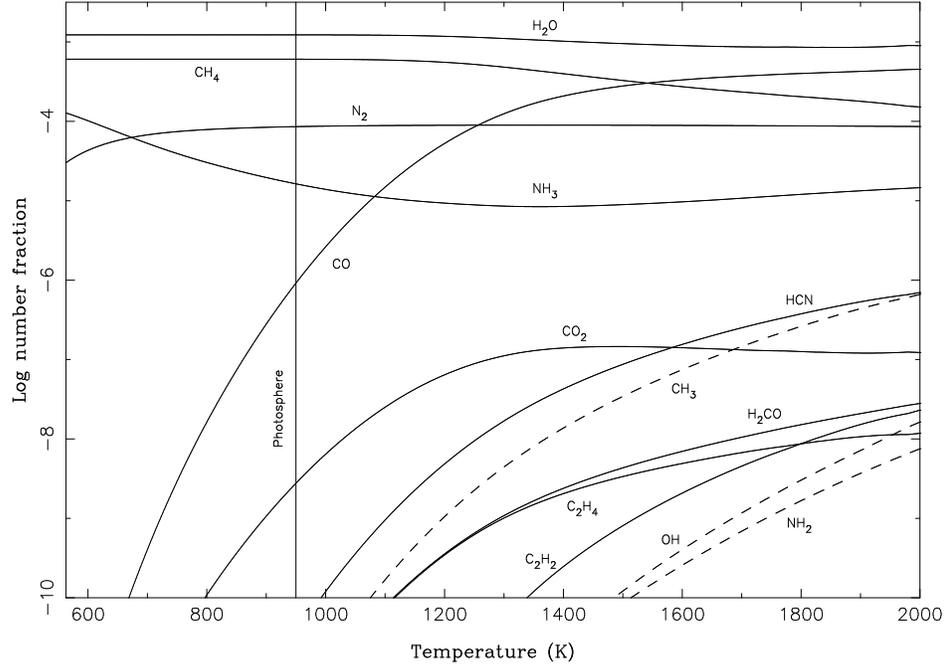}{3truein}{-90}{50}{50}{-200}{300}
\caption{Mixing fraction profiles in the gas phase as a function of temperature of the
major species, excluding H$_2$ and He, for a Gliese 229B model with an effective
temperature of 950 K and surface gravity of 1000 m/s$^2$.  Each curve shows the
equilibrium abundance of the indicated species, with pressure
changing with temperature according to the model.  The vertical line at 950 K marks the
photosphere, with the interior of the model being to the right.
The species that are chemically stable are shown as solid curves, with the radicals
CH$_3$, NH$_2$ and OH shown as dashed curves.  Of particular note is that as the temperature
decreases the dominant form of carbon changes from CO to CH$_4$ and the dominant
form of nitrogen changes from N$_2$ to NH$_3$.} 
\end{figure}

\subsection{Chemical Equilibrium, Condensation, and Clouds}
 
In the Burrows \etal\ (1997) exploration of EGPs and brown dwarfs, the chemical
equilibrium calculations were performed with the ATLAS code and
data from Kurucz (1970).
Recently, to the ATLAS code and
data from Kurucz (1970), we have added the JANAF tables and C. Sharp's 
free--energy minimization code (150--1500 species).
Condensation of CH$_4$, NH$_3$, H$_2$O, Fe, and
MgSiO$_3$ was included using data from various sources, including Eisenberg \& Kauzmann (1969), the
Handbook of Chemistry and Physics (1993), and Lange's Handbook of Chemistry
(1979). 

For T$_{\rm eff}$s below 1300 K, we assumed that Al, Ca,
Ti and V were removed either by condensation or were dissolved in silicate
grains at about the MgSiO$_3$ condensation temperature.
These atoms are important because they lead to molecules that are strong
light absorbers, such as TiO and VO. However, they have not been
detected in the giant planets of our solar system and shouldn't be
present in the  atmospheres of relatively cool objects such as the brown dwarf Gl229 B
(Marley \etal\ 1996). Figure 5 depicts composition profiles of some of the dominant molecular
species found in one of our models of Gl229 B.  It is clear from just this representative figure that
the variation of composition is an important feature of dense, low--temperature atmospheres 
(Sharp and Wasserburg 1995).
We are currently preparing a more
comprehensive set of condensation curves for minor species
with the EGP and brown dwarf atmosphere models of Burrows \etal\ (1997).

The presence of condensed species can radically
alter the gas phase composition. For example, the abundance of H$_2$S decreases rapidly with
temperature due to the formation of the condensate FeS.
In our theoretical model of Gl229 B,  FeS, and of
lesser importance due to the relatively low cosmic abundance of phosphorus,
Mg$_3$P$_2$O$_8$, form above the photosphere.  This suggests that clouds
containing grains of these species can form in the atmosphere, and possibly
play an important role in modifying the emergent flux and in altering the Bond albedo.  The more refractory
condensates whose condensation points lie well below the photosphere still
play an important role in depleting the observed atmosphere of a number of
abundant elements, {\it e.g.},  Si, Mg, Ca, and Fe.
The direct effect of clouds on the emergent fluxes of EGPs and brown dwarfs in the important T$_{\rm eff}$ 
range below 1300 K
has yet to be properly addressed.
However, the presence or absence of clouds strongly affects the reflection albedos of EGPs and brown
dwarfs. 
 
Cloud formation depletes a gas--phase
absorber from certain regions of the atmosphere; if this occurs
around the photosphere the resulting radiative balance and emergent
flux distribution are modified.  Because of condensation, we expect that
the gaseous water bands will disappear for objects with effective
temperature below about 400 K and that ammonia bands will disappear below T$_{\rm eff}$s of 200 K.
We expect the disappearance of silicate or iron features below about
1200 K (depending, modestly, on surface gravity).

Beyond predicting where the water and ammonia bands should
disappear due to condensation, the spectral and radiative effects
of clouds are difficult to quantify. Simple models in which clouds
are uniformly distributed over the surface of the EGP, and are
characterized by a single particle size, fail to take account of
atmospheric dynamics which can lead to dramatic changes in the
effects of clouds. In particular, convective processes lead to
growth in the mean particle size, as well as a potentially
heterogenous distribution of clouds across the disk of the object.
In the case of water and magnesium silicates, the latent heat of
condensation increases the mean upwelling velocity and can
exaggerate these effects, as quantified by Lunine \etal\ (1989).
The simple model of the transport processes in magnesium silicate
clouds presented in Lunine \etal\ suggests particle sizes in the range of
100 microns are possible by coalescence, much larger than the
micron--sized particles one would assume from simple condensation. The
radiative properties of a cloud clearly depend upon the
actual particle size, as well as the large--scale cloud
morphology (broken or continuous).
 
It is now clear that silicate grain formation modifies the characteristics of hot brown dwarfs
($1200\ {\rm K} < {\rm T}_{\rm eff} < 3000\ {\rm K}$)
and late M dwarfs ($\sim 1750\ {\rm K} < {\rm T}_{\rm eff} < 3000\ {\rm K}$) in ways not fully understood 
(Lunine \etal\ 1989; Allard \etal\ 1997;
Jones \& Tsuji 1997; Tsuji \etal\ 1996).  In particular, the near--infrared and optical colors of the objects 
newly discovered in the
DENIS survey (Delfosse 1997), the 2MASS survey (Kirkpatrick, Beichman, \& Skrutskie 1997),
and in the Pleiades (Martin, Rebolo, \& Zapatero-Osorio 1996) deviate
from what is currently theoretically predicted by as much as
one magnitude.  Our group was the first to include silicate grains in calculations of the
atmospheres of M and brown dwarfs (Lunine \etal\ 1986; Lunine \etal\ 1989).  However,
we produced only very low resolution spectra, and those were solely for the purposes of calculating boundary 
conditions
for the evolutionary calculations, not for the generation of colors and useful spectra for objects above 2000 K.
One of our goals for the future is to model the effects of refractory clouds at high
T$_{\rm eff}$s and H$_2$O and NH$_3$ clouds at low T$_{\rm eff}$s on emergent spectra and colors.

\section{Conclusions}

Our previous theoretical work has led us to certain general conclusions, among which are:
 
\medskip

\begin{itemize}

\item The opacity due to grains lowers the T$_{\rm eff}$, the luminosity, and the mass
at the edge of the solar--metallicity hydrogen--burning main sequence from $\sim$2000 K to 1750 K,
from 10$^{-4}$\lo to 6.0$\times$10$^{-5}$\lo, and from $\sim$0.085\ \mo to $\sim$0.074\ \mo, respectively.
 
\item  H$_2$O, H$_2$, and CH$_4$ dominate the spectrum below T$_{\rm eff}$$\sim$1200 K.
For such T$_{\rm eff}$s, most or all true metals are sequestered below the photosphere.
 
\item Enhancements and suppressions of the emergent flux relative to black body values can be by
many orders of magnitude.
 
\item The infrared colors of EGPs, brown dwarfs, and M dwarfs are much bluer than the colors previously
derived using either the black body assumption or primitive non--gray models.
 
\item In some IR colors ({\it e.g.,} $J-K$), a substellar object gets bluer, not redder, with age
and for a given age, lower--mass substellar objects are bluer than higher--mass substellar objects.
 
\item Clouds of H$_2$O and NH$_3$ are formed for T$_{\rm eff}$s
below $\sim$400 K and $\sim$200 K, respectively.  Their formation will affect the colors and spectra
of EGPs and brown dwarfs in ways not yet fully characterized.

\end{itemize}

\medskip

The present pace of M dwarf, brown dwarf, and giant planet discovery and NASA's future plans
for planet searches suggest that
many more objects with T$_{\rm eff}$s from 200 K to 4000 K
will be identified and subject to spectroscopic examination. 
We have now developed most of the theoretical tools necessary to support and interpret
indirect observations (radial velocity, astrometry), direct detections, and the
spectroscopy of the brighter objects.
We will soon be able to reliably predict the spectral and broadband signatures
of those objects that have not been directly detected, and constrain the
atmospheric composition and thermal structure of those for which we do have color or
spectroscopic information.

It is important to note that there is still considerable room for improvement 
in the modeling of substellar objects.  The direct effects of grains and clouds
for T$_{\rm eff}$s from 2000 K to 100 K has yet to be properly included and, in particular,
should mute the extreme suppressions we see due to the H$_2$O bands.  In our recent work, we
generated low--resolution spectra; higher--resolution
spectra are necessary to obtain more accurate colors (Figure 4).  The effect of
metallicity has yet to be explored and promises to bring many surprises.
Some opacity databases, especially that for CH$_4$, are inadequate and bear directly
upon the detectability and colors of EGPs.  All in all, giant planet and brown dwarf
theory should be a quite lively arena in the years ahead.

\acknowledgments

We thank F. Allard, W. Benz, S. Kulkarni, J. Liebert, B. MacIntosh, G. Marcy, M. Mayor, and B. Oppenheimer
for useful contributions and conversations.
This work was supported under NSF grants AST-9318970 and AST-9624878, under
NASA grants NAG5-2817, NAGW-2250, and NAG2-6007, and 
in part by the NSF under grant PHY94-07194 to the Institute for Theoretical Physics at Santa Barbara,
California.

\begin{question}{Dr. John Stauffer}
Are there specific gravity--dependent differences predicted in the
spectra between a 70 Myr old 0.03 \mo or 0.04 \mo brown dwarf and 
a VLM stellar object of the same luminosity and T$_{\rm eff}$
(other than Lithium)? 
\end{question}
\begin{answer}{Adam Burrows}
At a Pleiades age, T$_{\rm eff}$ for 0.03 \mo or 0.04 \mo brown dwarfs is from 500 to 1000 K cooler
than for 0.08 \mo or 0.1 \mo stars.  The temperature differences alone will be
sufficient to discriminate stellar from substellar, as will grain effects. 
In $J-H$, $J-K$, $H-K$,  $K-L^{\prime}$, $M-N$, the IR color differences between such brown dwarfs and a 
late M dwarf should 
be at least 0.5 magnitudes, with the star being at times bluer, at times redder, depending upon the
color.  In addition, the luminosity of such a 70 Myr brown dwarf is an order of magnitude
lower than that of a coeval object destined to ignite hydrogen stably.
However, the gravity differences per se are not large, and translate into perhaps a few tenths of a magnitude;
in this, theory is not yet precise enough to pontificate.
Work aplenty remains.
In order to obtain sufficiently good gravities, I surmise 
that modest--resolution ($\lambda/\Delta\lambda \sim 200$)
IR spectroscopy and spectrophotometry will be necessary.  
\end{answer}

\begin{question}{Dr. Axel Brandenburg}
You mentioned using mixing--length theory, but then you
didn't talk about possible convection zones.  Could you please comment on the existence,
location, and nature of convective zones in brown dwarfs?  What are typical convective
velocities?  Do brown dwarfs have magnetic fields?
\end{question}
\begin{answer}{Adam Burrows}
This is a large subject in itself and has received extensive comment. 
In sum, brown dwarfs in isolation
are almost completely convective; the outer radiative zones might comprise only 10$^{-10}$ \% by mass
of the object.  However, as Guillot \etal\ (1995) have shown, Jupiter itself may have an
inner radiative zone near 1500 K that can confuse the entropy mapping between
core and atmosphere.  We see the same effect in brown dwarfs and I refer you to our
recent paper (Burrows \etal\ 1997) for a more detailed discussion of this important point.
Being convective, brown dwarfs should have magnetic fields and be active, though
their luminosties are low and their convective velocities are correspondingly low
($v \propto L^{1/3}$). 
\end{answer}

\begin{question}{Dr. Mark Giampapa}
For potential searches for brown dwarfs/large planets
using photometric transits, the magnitude of the effect of warming on the radius
of these objects which are in proximity to the primary becomes important.
What is the magnitude of this effect?
\end{question}
\begin{answer}{Adam Burrows}
For the companion to 51 Peg, we have estimated this effect to be a mere $\sim$20\%.
At a given distance, in a steady--state with stellar insolation lower--mass objects are generally larger. 
Younger objects may still be contracting on their Hayashi tracks and, hence,
can also be larger than after they achieve equilibrium, but the radii of objects
greater than one Jupiter mass are less than 2 \rj within 10$^6$ yrs.
There are some interesting exceptions, but these are good general rules;  the effect is
generally small.
\end{answer}

\begin{question}{Dr. B.H. Foing}
You showed the high potential of IR searches for isolated brown dwarfs.
What strategies do you recommend for brown dwarfs companions, in order to
discriminate them spatially or spectroscopically from the primary star?
\end{question}
\begin{answer}{Adam Burrows}
The best long--term means of discriminating a brown dwarf or a planet from
its primary are adaptive optics and interferometry in the near-- to mid--IR,
where the contrast is greatest and the Fried parameter is advantageous.
Another potentially rewarding technique is to look for the contrast
across a feature, such as one of methane's, unique to substellar objects.  This 
approach has already been demonstrated for Gl229 B by Rosenthal \etal.
\end{answer}

\end{document}